\documentclass[11pt,a4paper]{article}
\pdfoutput=1

\usepackage{graphicx}
\usepackage{dcolumn}
\usepackage{bm}
\usepackage{amsthm}
\usepackage{amsmath}
\usepackage{color}

\usepackage{jheppub}

\title{Spontaneous Breaking of Non-Relativistic Scale Symmetry}

\author{Igal Arav,}
\author{Itamar Hason,}
\author{Yaron Oz}
\affiliation{Raymond and Beverly Sackler School of Physics and Astronomy, Tel Aviv University, Tel Aviv 69978, Israel}
\emailAdd{aravigal@post.tau.ac.il}
\emailAdd{itamarhason@gmail.com}
\emailAdd{yaronoz@post.tau.ac.il}

\date{\today}

\abstract{
We analyze the mechanism of spontaneous symmetry breaking of scale invariance in Galilean invariant field theories. We show that the existence of a dynamic gapless dilaton mode depends on whether the $U(1)$ particle number or the Galilean boost symmetry are spontaneously broken. When both scale and particle number symmetries are spontaneously broken there is one propagating gapless Nambu-Goldstone mode. Its dispersion relation is linear if the chemical potential is nonzero and quadratic otherwise. We discuss the reversibility of RG flows in such theories.
}


\begin{document}
\maketitle

\section{Introduction}

Spontaneous symmetry breaking (SSB) underlies a large number of physical phenomena such as superconductivity, superfluidity and the generation of elementary particle masses. A spontaneous breaking of a continuous global symmetry  implies via the Nambu-Goldstone (NG) theorem the existence of a gapless  Nambu-Goldstone mode.
In a relativistic field theory there is one NG mode for each broken symmetry generator \cite{Nambu:1960,Goldstone:1961,Goldstone:1962}. This one to one correspondence between broken generators and gapless NG modes does not hold when spacetime symmetries are spontaneously broken \cite{Low:2001bw,Watanabe:2013iia}. It is also not generally the case for spontaneously broken global symmetries in non-relativistic field theories \cite{Nielsen:1975hm, Watanabe:2012hr, Nicolis:2012vf}.

When scale symmetry is spontaneously broken in a relativistic field theory, there is a corresponding NG mode, the dilaton. The dilaton effective action encodes, for instance, the information about the A-type conformal anomaly and has been valuable in proving the a-theorem  in \cite{Komargodski:2011vj}. This leads naturally to inquire about the mechanism of spontaneously broken scale symmetry in non-relativistic field theories. Such field theories have much importance in the study of low energy condensed matter systems, as well as in non-relativistic holography (e.g. \cite{Son:2008ye}). 

The aim of this paper is to analyze SSB of scale invariance in Galilean invariant field theories. In non-relativistic field theories space and time scale differently: $\vec{x} \rightarrow e^{\sigma}  \vec{x}, t \rightarrow e^{z\sigma}  t$, where  $\sigma$ is a real parameter and $\vec{x}$ is a $d$-dimensional vector. $z$ is called a dynamical exponent and we will consider the Galilean case $z=2$. We will show that there is no gapless dilaton mode in a Galilean field theory unless the $U(1)$ particle number symmetry or the Galilean boost symmetry are also spontaneously broken. Particle number symmetry is an additional symmetry of  Galilean field theories that does not exist in the relativistic case. We will see that when both scale and particle number symmetries are spontaneously broken there is just one propagating gapless mode.\footnote{The case where the particle number symmetry is not broken but the Galilean boosts are broken seems to be of less physical relevance and will be discussed in the appendix.}

In the Galilean algebra the generator $M$ of the $U(1)$ particle number symmetry is a central extension. It appears in the commutator of  translations $P_i$ and Galilean boosts $K_j$
\begin{equation}
[P_i, K_j] = -i\delta_{ij}M,~~~~i=1,...,d \ .
\label{com}
\end{equation}

The broken symmetry generators create NG modes from the vacuum and their commutator algebra may impose relations between these modes. Such an example is the relation among the NG mode related to SSB of the $U(1)$ particle number symmetry and the modes related to the SSB of boosts. While commutator \eqref{com} predicts that $U(1)$ SSB implies that Galilean boosts are spontaneously broken, it also predicts that there is only one independent NG mode. This relationship is an example of a general structure called Inverse Higgs Constraints (IHC) \cite{Ivanov:1975zq,Brauner:2014aha,Nicolis:2013sga}. Such an argument, however, does not explain why when scale and $U(1)$ are spontaneously broken there is only one NG mode. We will derive this result by an explicit calculation of the spectrum.
 
Based on $z=2$ dimensional arguments we may anticipate the dispersion relation of the NG mode. If we denote by $v$ the symmetry breaking length scale, then the dispersion relation takes the general form
\begin{equation}
 \omega = \frac{\vec{k}^2}{2m} F\left({kv}\right) \ .
\end{equation}
We'll see that this is indeed the behavior, where with zero chemical potential $F(0)$ is finite and we get to leading order a quadratic dispersion relation while with a nonzero chemical potential we get to leading order a linear dispersion relation instead of a quadratic one.

There are various methods to construct the effective action of NG bosons. One way is to write all possible terms that respect all the symmetries. Another way is to couple the theory to curved external background sources. We will consider both methods in the study of the non-relativistic dilaton.\footnote{Another possible method utilizes the coset construction, however we will not pursue it in this work.}

The paper is organized as follows. In section \ref{sec:setup} we will outline the setup of the problem and the various cases that we consider. In the next section we will begin by considering the NG effective action based on symmetry arguments. We will see that spontaneously breaking scale invariance while maintaining the $U(1)$ and Galilean boost symmetries does not allow for a propagating dilaton mode. We will then spontaneously break the $U(1)$ and boost symmetries as well. We will construct the NG effective action at leading orders in the derivatives expansion. Next, we will derive the same effective action using a coupling to the Newton-Cartan curved geometry. We will analyze the spectrum of the NG bosons and find one gapless propagating mode. Finally, we will discuss the possible relevance of the results to RG flow theorems in Galilean field theories and conclude with a brief summary and outlook. In appendix \ref{boostsAppendix} we discuss the case of unbroken particle number and broken boost symmetries.

\section{The Problem Setup} \label{sec:setup}

We consider non-relativistic systems in $d+1$ dimensions, which are invariant under the (centrally extended) Galilean group with the addition of non-relativistic (Lifshitz) scale invariance with a dynamical exponent of $z=2$. The symmetry generators satisfy the following algebra:
\begin{equation}\label{SymAlgebra}
\begin{split}
&[L_{ij},L_{kl}] = i \left[ \delta_{ik} L_{jl} - \delta_{jk} L_{il} + \delta_{il} L_{kj} - \delta_{jl} L_{ki} \right], \\
&[L_{ij},P_k] = i \left[ \delta_{ik} P_j - \delta_{jk} P_i \right], \qquad\qquad
[L_{ij},K_k] = i \left[ \delta_{ik} K_j - \delta_{jk} K_i \right], \\
&[K_i,H] = i P_i, \qquad\qquad\qquad\qquad\qquad
[K_i, P_j] = i \delta_{ij} M, \\
&[D,H] = i 2H, \qquad
[D,P_i] = i P_i,  \qquad
[D,K_i] = -i K_i,
\end{split}
\end{equation}
where $H$, $P_i$, $L_{ij}$, $K_i$ and $D$ are the generators of time translations, space translations, space rotations, Galilean boosts and Lifshitz scaling respectively. $M$ is a generator of the $U(1)$ internal symmetry that corresponds to the conserved particle number, and all other commutators are zero. 
We will also discuss systems which are invariant under the full Schr\"{o}dinger group, which contains, in addition to these symmetries, the special conformal transformations, the generator of which we denote by $C$. In addition to the above algebra \eqref{SymAlgebra}, the generators then also satisfy:
\begin{equation}\label{SpecialConfAlgebra}
[C,P_i] = -iK_i, \qquad
[C,H] = -iD, \qquad
[D,C] = -i 2C .
\end{equation}
We assume either that the system is in its vacuum state that minimizes the Hamiltonian $H$, or alternatively that the state of the system minimizes the modified Hamiltonian $\tilde{H} \equiv H - \mu M $, where $\mu$ is the chemical potential of the system.\footnote{For describing a system with a non-vanishing chemical potential $\mu\neq 0$ we take the point of view of e.g.\ \cite{Nicolis:2013sga, Endlich:2013spa}, that the state of the system minimizes $\tilde{H}$, but the system still propagates in time according to the Hamiltonian $H$. Also note that it is $H$, rather than $\tilde{H}$, that appears in the algebra \eqref{SymAlgebra}.}

Such Galilean systems are usually observed in nature simply as low velocity limits of relativistic systems. In these cases, the corresponding relativistic systems have a relativistic Hamiltonian $H_\text{rel}$ and relativistic chemical potential $\mu_\text{rel}$ associated with them, and one defines the non-relativistic Hamiltonian and chemical potential as:
\begin{align}
H &\equiv H_\text{rel} - M c^2 , \\
\mu &\equiv \mu_\text{rel} - c^2 ,
\end{align}
and takes the limit of $ v \ll c$ (or $ p c \ll E$). It is important to note that, even in such cases, the Hamiltonian $H$ and chemical potential $\mu$ we refer to in the following sections are the \emph{non-relativistic} ones, and it is the non-relativistic Hamiltonian $H$ that satisfies the above algebra. It is also worth mentioning that, while such a Galilean symmetry often appears as the small velocity limit of a full Lorentzian relativistic symmetry, it may also appear as an emergent symmetry at low energy.

In physical systems, the particle number $M$ generally has a non-negative spectrum, where the only state with $M=0$ is the empty state that contains no particles. The case of $\langle M \rangle =0$ (with zero particle number density $\rho$) is therefore not physically interesting for the study of SSB, as we don't expect any symmetry to be spontaneously broken in such cases. However, in the interest of mathematical completeness (and since, as we point out later, the study of the dilaton effective action is useful for analyzing other QFT properties outside of SSB), we will briefly consider this case. The more physically interesting case, however, is that for which $\langle M \rangle > 0$. In this case, the Galilean boost symmetries are necessarily spontaneously broken by the particle number current $J_i(x)$ since $\langle [K_j,J_i(x)] \rangle \propto \delta_{ij} \langle \rho(x) \rangle \neq 0 $. In both cases, for the purposes of this work, we will assume that space translations and space rotation symmetries (as well as the modified time translation generated by $\tilde{H}$) are not broken. 

We will consider systems for which the non-relativistic scale symmetry $D$ is spontaneously broken, that is, there is some dimensionful operator $ \mathcal{O} $ such that $ \langle [D,\mathcal{O}] \rangle \propto \langle \mathcal{O} \rangle \neq 0 $. There are then two possible cases depending on whether the operator $\mathcal{O}$ is invariant under the particle number $U(1)$ symmetry.

If the operator $\mathcal{O}$ is $U(1)$ invariant (and assuming all other operators with non vanishing expectation values are as well), the $U(1)$ symmetry is not spontaneously broken. Assuming non-vanishing particle number $ \langle M \rangle \neq 0 $ \footnote{Note that the condition $\langle M \rangle \neq 0 $ does not necessarily imply that the $U(1)$ symmetry is spontaneously broken (which requires the existence of an operator $\mathcal{O}$ such that $\langle [M,\mathcal{O}] \rangle \neq 0 $). If the system is in any eigenstate of $M$, for example, then $M$ cannot be spontaneously broken, as pointed out in \cite{Endlich:2013spa}.}, this is a case where only the scale and Galilean boost symmetries are broken. Such a system is similar to the ``framids'' discussed in \cite{Nicolis:2015sra}, in the non-relativistic limit. One might expect the appearance of $d+1$ NG modes, corresponding to the broken scale and Galilean boosts. Since it is not clear whether such a scenario indeed appears in any physical system, we relegate this case to appendix \ref{boostsAppendix}.

If the operator $\mathcal{O}$ is not $U(1)$ invariant, and $\langle [M,\mathcal{O}] \rangle \neq 0 $, the $U(1)$ symmetry is spontaneously broken as well. Examples of such systems may be found in superfluids, superconductors and other types of non-relativistic condensates. In these systems, since both the scale, the boosts and the $U(1)$ symmetries are spontaneously broken, one might initially expect to find separate NG modes for each of them. However, as explained in e.g.\ \cite{Brauner:2014aha}, this is not the case: Due to the well-known ``Inverse Higgs'' mechanism, the commutation relation between $K_i$ and $P_j$ in \eqref{SymAlgebra} imposes a relation between the modes $v_i$ that would otherwise correspond to the broken Galilean boosts and the mode $\theta$ corresponding to the broken $U(1)$ symmetry, of the form: $ v_i \propto \partial_i \theta $. Therefore, in this case, the \emph{broken} Galilean boosts play no role when building the possible effective action for the NG modes -- one need only consider the modes corresponding to the broken scale and $U(1)$ symmetries (while the effective action is, of course, required to be invariant under all of these symmetries, including Galilean boosts).

For systems which are Schr\"{o}dinger invariant, it is clear from the relations \eqref{SpecialConfAlgebra} that, in all the cases considered, the special conformal symmetry $C$ is necessarily also broken. However, again due to the ``Inverse Higgs'' constraints derived from \eqref{SpecialConfAlgebra} (and similarly to the relativistic conformal case), we do not expect a corresponding independent NG mode to appear as the algebra imposes a relation between this mode and the other ones.

When considering the case of broken $U(1)$, it is important to distinguish between two possible sub-cases, corresponding to the existence of a non-vanishing chemical potential $\mu$. When $\mu=0$, the expectation of the operator $\mathcal{O}$ is constant in time, since in the vacuum state:
\begin{equation}
\partial_t \langle \mathcal{O} \rangle = -i \langle [H,\mathcal{O}] \rangle = 0.
\end{equation}
However, as pointed out in \cite{Endlich:2013spa}, when $\mu \neq 0$, this is not the case and one instead obtains:
\begin{equation}
\partial_t \langle \mathcal{O} \rangle = -i \langle [H,\mathcal{O}] \rangle = 
-i \mu \langle [M,\mathcal{O}] \rangle = -i m \mu \langle \mathcal{O} \rangle
\quad \Rightarrow \quad
\langle \mathcal{O} \rangle \propto e^{-i m \mu t},
\end{equation}
assuming that $\mathcal{O}$ has a charge of $m$ under the $U(1)$ symmetry. This is accounted for in our spectrum analysis in subsection \ref{subsec:EffActionBrokenU1} by perturbing the $U(1)$ NG mode $\theta$ around $-\mu t$.

Note that generally, the condition of non-vanishing particle number $\langle M \rangle \neq 0$ doesn't necessarily imply a non-vanishing chemical potential. As a simple example, consider a model for a mixture of two Bose condensates in $2+1$ dimensions with contact interactions, given by the following action (this is a special case of the model considered e.g.\ in \cite{Lellouch:2013}):
\begin{equation}
S = \int dt d^2 x \left[ \sum_{k=1}^2 \left( i \phi_k^* \partial_t \phi_k - \frac{1}{2m} \partial_i \phi_k^* \partial_i \phi_k \right)
- g \left( |\phi_1|^2 - |\phi_2|^2 \right)^2 \right],
\end{equation}
where $m$ is the Galilean mass of the condensates, and $g$ a (dimensionless) coupling constant.  
This model is indeed Galilean invariant, and classically it is Lifshitz scale invariant as well. With vanishing (non-relativisitic) chemical potential, this model still classically allows for a moduli space of (inequivalent) vacua with a finite particle number density, where both $\phi_1$ and $\phi_2$ gain a non-vanishing expectation value such that $ |\phi_1|^2 = |\phi_2|^2 \neq 0$. Each of these vacua breaks the $U(1)$, boosts and scale symmetry spontaneously. While this is only true without considering any quantum corrections, this example serves to demonstrate the possibility of having a finite particle density configuration with vanishing chemical potential.

To summarize, in this work we consider the following cases:
\begin{itemize}
\item The case where only scale symmetry is spontaneously broken (the non-physical $M=0$ case),
\item The case where both scale and Galilean boost symmetries are broken but the $U(1)$ particle number symmetry remains unbroken (considered in appendix \ref{boostsAppendix}),
\item The case where scale, Galilean boosts and the $U(1)$ symmetry are all spontaneously broken, either with vanishing or finite chemical potential.
\end{itemize}
Our main focus is on the third case, which seems relevant to various condensed matter systems. For each case, we will derive the possible form of the effective action for the NG modes, and study their spectrum.
 
Finally, it is important to note that the analysis of effective actions is also useful for the study of other QFT properties other than SSB. In relativistic theories, for example, it has been used to study quantum anomalies and RG flow properties of field theories (see \cite{Komargodski:2011vj}). As such, even the cases which are less physically relevant in the context of SSB may be helpful for drawing conclusions on other properties of non-relativistic QFTs. In section \ref{sec:atheorem} we indeed comment on the possibility of applying some of our conclusions to the study of RG flows in Galilean field theories.

\section{The NG Boson Effective Action}

\subsection{Symmetries}

In this subsection we derive the first few terms in a derivative expansion of the effective action of the non-relativistic NG boson following a spontaneous breaking of scale invariance $(\vec{x},t) \rightarrow (e^{\sigma} \vec{x},  e^{z\sigma}t )$, using the symmetries of the theory.

\subsubsection{Unbroken $U(1)$ Particle Number and Galilean Boost Symmetries}

The dilaton is a real field which we will denote by $\tau$. It carries zero $U(1)$ charge. Under scale transformation the dilaton transforms as
\begin{equation}
 \tau \left( {\vec{x},t} \right) \rightarrow \tau\left({e^{\sigma} \vec{x},  e^{z\sigma}t}\right) + \sigma \ .
\end{equation}
Since $dt d^{d}x \rightarrow e^{-\left({ z+d } \right) \sigma }dtd^{d}x$, $\partial_t \rightarrow e^{z\sigma}\partial_t$ and 
$\partial_i \rightarrow e^{\sigma} \partial_i$, the effective Lagrangian density should take the form 
\begin{equation}
 L = {\cal L} \left( { e^{-z\tau}\partial_t, e^{-\tau} \partial_i} \right) e^{\left({z+d}\right)\tau} \ ,
\end{equation}
where $\cal L$ is a general polynomial in these differential operators.

Under $z=2$ non-relativistic boost transformation with boost parameters $\vec{u}$,  the dilaton and its derivatives transform as
\begin{equation}
\begin{split}
&\tau\left({\vec{x},t}\right) \rightarrow \tau \left({\vec{x}-\vec{u}t, t}\right), \\
&\partial_i\tau\left({\vec{x},t}\right) \rightarrow \partial_i\tau\left({\vec{x}-\vec{u}t,t}\right), \\
&\partial_t\tau\left({\vec{x},t}\right) \rightarrow \partial_t\tau\left({\vec{x}-\vec{u}t,t}\right) - u_i\partial_i\tau\left({\vec{x}-\vec{u}t,t}\right) \ .
\end{split}
\end{equation}
The $\partial_i$ term is invariant while the $\partial_t$ term is not. This basically forbids any time derivative term in the effective Lagrangian, hence there is no dynamical dilaton.

Consider this more explicitly.
Zeroth order derivative terms can't by themselves generate dynamics, however, they are allowed provided the scale transformation is fixed correctly as follows:
\begin{equation} \label{L0}
L_0 = \Lambda e^{\left(z+d\right)\tau}  \ ,
\end{equation}
where $\Lambda$ is a dimensionless constant.
First order derivatives terms are total derivatives, so next we consider the second order derivatives terms. By rotational symmetry, the only possible
two derivatives terms are $\left( { \partial_t \tau } \right) ^ 2$ and $\left( { \partial_i \tau } \right) ^ 2$.
Thus, up to two derivatives the effective dilaton Lagrangian should read
\begin{equation}
 L = e^{\left({-z+d}\right)\tau} \left( { \partial_t \tau } \right) ^ 2 - \gamma e^{\left({z+d-2}\right)\tau} \left( { \partial_i \tau } \right) ^ 2 + \Lambda e^{\left(z+d\right)\tau} \ .
\end{equation}
The first term is not invariant under $z=2$ non-relativistic boost transformation.
One can repeat the same analysis at higher order in derivatives arriving to the same conclusion that there is no dynamical dilaton. Note, that if one works in Lifshitz field theory but without imposing Galilean boost invariance, one can have a dynamical dilaton. For example, the $z=2$ scale invariant terms 
$e^{\left({d-2}\right)\tau} \left( { \partial_t \tau } \right) ^ 2, e^{\left({d-2}\right) \tau} \left({\partial_i^2 \tau}\right)^2, e^{\left({d-2}\right) \tau} \left({\partial_i \tau}\right)^4, e^{\left({d-2}\right) \tau} \left( \partial_i \tau \right)^2 \left( \partial_i^2 \tau \right)$, combine using 
$\phi = e^{ \frac{\left(d-2\right)}{2}\tau}$ to give 
\begin{equation}
 L = \frac{1}{2}\left( { \partial_t \phi } \right) ^2  -\frac{\kappa}{4} \left( { \partial_i^2 \phi } \right) ^2 \ .
\end{equation}
This is the Lagrangian for a free Lifshitz real scalar, which is not boost invariant.

\subsubsection{Broken $U(1)$ Particle Number Symmetry}
\label{subsec:EffActionBrokenU1}

In the following both scale and $U(1)$ particle number are spontaneously broken and we will consider a complex scalar field $\phi = e^{\Delta\tau+im\theta}$ that carries a $U(1)$ charge $m$ and scaling dimension $\Delta$ (where $\tau$ corresponds to the dilaton and $\theta$ to the $U(1)$ NG mode). As explained above, Galilean boosts must also be broken but the corresponding NG modes are directly related to the $U(1)$ NG mode, and are therefore not independent modes.\footnote{Note that, since the boost NG modes are related to the $U(1)$ mode by a simple space derivative, i.e.\ $v_i = \partial_i \theta$, any effective action terms allowed by the symmetries that can be constructed from the boost NG modes are included in the types of Lagrangian considered here, built from $\phi$. This can be seen by expressing these Lagrangians in terms of $\theta$ and $\tau$, expanding them in powers of $\Delta$ and $m$ and extracting the various independent terms as the expansion  coefficients (including all the ones that contain only $v_i, \tau$ and their derivatives).
For example, the $B$ term in \eqref{boostAppendixEffectiveAction} can be expressed as \eqref{eq:L2detailed} with $\Delta=0$ and $m \to 0$.}

Under $z=2$ boost transformation one has
\begin{equation}
\phi \left({\vec{x},t}\right) \rightarrow \phi \left( {\vec{x}-\vec{u}t,t}\right)e^{-\frac{i}{2}m\vec{u}^2t+im\vec{u}\cdot\vec{x}} \ ,
\end{equation}
or $\tau$ invariant and
\begin{equation} \label{eq:theta_boost_trans}
 \theta \rightarrow \theta-\frac{1}{2}\vec{u}^2t+\vec{u}\cdot\vec{x} \ .
\end{equation}
The zero derivatives term $\phi\phi^*$ is invariant and gives \eqref{L0} with $\Delta=\frac{d+2}{2}$.
At leading order in derivatives we can write a scale, rotation and boost invariant Lagrangian
\begin{equation}
L_1 = \frac{i}{2} \left( { \phi^*\partial_t\phi-\partial_t\phi^*\phi } \right) - \frac{1}{2m}{\partial_i\phi}{\partial_i\phi^*} \ ,
\label{L1}
\end{equation}
with $\Delta=\frac{d}{2}$.

More generally, we can build a boost invariant structure (up to a phase)
$-i\partial_t\phi - \frac{1}{2m}\partial_i^2\phi$.
It can be used for the construction of a boost and scale invariant Lagrangian, so to next order in derivatives one has also
\begin{equation}
L_2 = \left( { i\partial_t\phi^* - \frac{1}{2m}\partial_i^2\phi^* } \right) 
\left( { -i\partial_t\phi -\frac{1}{2m}\partial_i^2\phi } \right)  e^{2\left(\Delta'-\Delta\right)\tau} \  ,
\label{L2}
\end{equation}
where $\phi$ has $U(1)$ charge $m$ and scaling dimension $\Delta$, and $\Delta' = \frac{d}{2}-1$.
In terms of $\tau$ and $\theta$
\begin{equation}
L_2 =  \left|  { i\partial_t\left( { \Delta\tau -i m\theta } \right) - \frac{1}{2m} \left( { \partial_i\left( { \Delta\tau - im\theta } \right)} \right) ^ 2 - \frac{1}{2m} \partial_i^2 \left( { \Delta\tau - im\theta } \right)}  \right| ^ 2 e^{2\Delta'\tau} .  \ 
\label{eq:L2detailed}
\end{equation}
Of course, we may similarly construct other invariant terms of this order in derivatives (or higher) using the field $\phi$, the covariant derivative operator $-i\partial_t - \frac{1}{2m}\partial_i^2$ and appropriate exponents of $\tau$ to compensate for the dimension. Note that in subsection \ref{subsec:BrokenU1Spectrum} we use a term of the form \eqref{L2} as an example of an higher order term for the purpose of analyzing the spectrum of the theory, however using other higher order terms constructed this way does not change the main results of that subsection.

We can also consider the case of theories which are invariant under the full Schr\"{o}dinger group, which contains in addition the special conformal transformation, given by: 
\begin{equation}
\phi \left( \vec{x},t \right) \rightarrow \left( 1+\nu t \right)^{-\Delta} \, e^{i \frac{m}{2}\frac{\nu x^2}{1+\nu t}} \,\phi \left( \frac{\vec{x}}{1+\nu t}, \frac{t}{1+\nu t} \right),
\end{equation}
or in terms of $\tau$ and $\theta$:
\begin{equation}\label{eq:SpecialConformalTrans}
\begin{split}
\tau \left( \vec{x},t \right) &\rightarrow \tau \left( \frac{\vec{x}}{1+\nu t} , \frac{t}{1+\nu t} \right) - \ln (1+\nu t), \\
\theta \left( \vec{x},t \right) &\rightarrow \theta \left( \frac{\vec{x}}{1+\nu t} , \frac{t}{1+\nu t} \right) + \frac{\nu x^2}{2(1+\nu t)}.
\end{split}
\end{equation}
It is easily verified that the previously mentioned structure
$-i\partial_t\phi - \frac{1}{2m}\partial_i^2\phi$
is covariant under these special conformal transformations (that is, invariant up to a phase and a scale factor), only when $\Delta = \frac{d}{2}$ (or when $m=0$, which corresponds to terms that depend only on $\tau$ and its spatial derivatives). Therefore the Lagrangians \eqref{L0} (with $\Delta=\frac{d+2}{2} $) and \eqref{L1} (with $\Delta=\frac{d}{2}$) are indeed Schr\"{o}dinger invariant. However, the requirement for Schr\"{o}dinger invariance restricts the higher order Lagrangian \eqref{L2} to the case of $\Delta = \frac{d}{2}$ (or alternatively $m=0$).

\subsection{Geometry}\label{subsec:BrokenU1Geometry}
One can construct the NG boson effective action by geometrical considerations: coupling the field theory to a curved background, promoting the symmetries to local ones, looking at all possible actions in this framework and taking the flat background limit. This was done for the relativistic case \cite{Schwimmer:2010za} where the curved background is a Riemannian geometry with a metric tensor $G_{\mu\nu}$. One constructs a Weyl invariant metric $\hat{G}_{\mu\nu} =  e^{-2\tau} G_{\mu\nu}$ and writes the effective action in terms of scalar terms constructed from it. 

In the non-relativisitic case one has to use the Newton-Cartan (NC) geometry instead (see e.g. \cite{Jensen:2014aia,Arav:2016xjc}). The NC geometry is built from a time direction described by a 1-form $n_\mu$, a spatial metric $h^{\mu\nu}$ orthogonal to $n_\mu$ and a $U(1)$ gauge field $A_\mu$ which couples to the conserved particle number current. Further, one defines a vector $v^\mu$ that satisfies $v^\mu n_\mu = 1$ and induces a metric $h_{\mu\nu}$ satisfying
\begin{equation}
h_{\mu\nu}v^{\nu} = 0, ~~~ h_{\mu\rho}h^{\nu\rho} = P^\nu_\mu = \delta^\nu_\mu - v^\nu n_\mu \ .
\end{equation}
These definitions are not unique because we can redefine $v^\mu$ using an arbitrary vector $\psi_\nu$
\begin{equation}\label{eq:v_milne_trans}
v^\mu \rightarrow v^\mu +h^{\mu\nu}\psi_\nu  \ ,
\end{equation}
and redefine $h_{\mu\nu}$ correspondingly:
\begin{equation}
h_{\mu\nu} \rightarrow h_{\mu\nu} - \left( n_\mu P^\rho_\nu + n_\nu P^\rho_\mu \right) \psi_\rho + n_\mu n_\nu h^{\rho\sigma} \psi_\rho \psi_\sigma  \ , 
\end{equation}
together with the following redefinition of $A_\mu$:
\begin{equation} \label{eq:A_milne_trans}
A_\mu \rightarrow A_\mu + P^\nu_\mu \psi_\nu - \frac{1}{2}n_\mu h^{\nu\rho}\psi_\nu\psi_\rho  \ .
\end{equation}
These transformations are the Milne boosts and one can define Milne boost invariant objects, as follows:
\begin{equation} \label{eq:milne_invariant}
v_A^\mu = v^\mu - h^{\mu\nu}A_\nu, ~~~ g_{\mu\nu} = \left(h_A\right)_{\mu\nu} = h_{\mu\nu} + n_\mu A_\nu + n_\nu A_\mu \ .
\end{equation}

Since we are interested in the ($z=2$ anisotropic-)Weyl (scale) and $U(1)$ symmetries, we need the Weyl and $U(1)$ transformations of the NC objects. The $U(1)$ gauge symmetry transforms only $A_\mu$ out of the basic structures:
\begin{equation}
A_\mu \rightarrow A_\mu +\partial_\mu \Lambda \ ,
\end{equation}
while the Weyl symmetry transforms $n_\mu$ and $h_{\mu\nu}$:
\begin{equation}\label{eq:Weyl_trans}
h_{\mu\nu} \rightarrow e^{2\sigma} h_{\mu\nu}, ~~~ n_\mu \rightarrow e^{2\sigma} n_\mu \ .
\end{equation}

There are two equivalent ways to construct the NG boson effective action for the scale and $U(1)$ spontaneous breaking. One is to introduce spectator fields, $\tau$ for the scale symmetry and $\theta$ for the $U(1)$ symmetry, with $\tau$ transforming under scale like a dilaton $\tau \rightarrow \tau + \sigma$, and $\theta$ transforming under $U(1)$ as $\theta \rightarrow \theta + \Lambda$ where $A_\mu \rightarrow A_\mu + \partial_\mu \Lambda$ (and both invariant under Milne boosts). Using these spectator fields and the NC structures $h,n,A$, we can then define Weyl and gauge invariant geometric quantities $\hat{h}, \hat{n}, \hat{A}$ as follows:
\begin{equation}
\hat{h}_{\mu\nu} \equiv e^{-2\tau} h_{\mu\nu}, \quad 
\hat{n}_\mu \equiv e^{-2\tau} n_\mu, \quad
\hat{A}_\mu \equiv A_\mu - \partial_\mu \theta.
\end{equation}
From these structures, we can build boost invariant scalars.
This procedure is made straightforward by defining the boost invariant metric $\hat{g}$ and vector $\hat{v}_A^\mu$ as follows:
\begin{equation}
\begin{split}
\hat{v}_A^\mu &\equiv e^{2\tau} \left( v^\mu - h^{\mu\nu} \left( A_\nu - \partial_\nu \theta \right) \right), \\
\hat{g}_{\mu\nu} &\equiv e^{-2\tau} \left( { h_{\mu\nu} + n_\mu \left( { A_\nu - \partial_\nu \theta } \right) + n_\nu \left( { A_\mu - \partial_\mu \theta } \right) } \right) \ .
\end{split}
\end{equation}
Finally, we take the limit where the geometry is flat:
\begin{equation} \label{eq:flat_geometry}
h^{\mu\nu}\partial_\mu\partial_\nu=\delta^{ij}\partial_i\partial_j, \hspace{3mm} A=0, \hspace{3mm} n_\mu dx^\mu=dt, \hspace{3mm} v^\mu\partial_\mu = \partial_t ,
\end{equation}
and thus remain only with the spectator fields. 
This proposal is a generalization of the relativistic case discussed above, where the dilaton factor was added to the metric $G_{\mu\nu}$  to compensate for the Weyl variation. In our case we have the boost invariant structures $g_{\mu\nu}$ and $v_A^\mu$ given in  \eqref{eq:milne_invariant}, as well as $n_\mu$ and $h^{\mu\nu}$.
Here again, to enforce Weyl invariance we add $\tau$, and to enforce $U(1)$ invariance we add $\theta$. The structures $\hat{g}_{\mu\nu}, \hat{v}_A^\mu, \hat{n}_\mu, \hat{h}^{\mu\nu}$ are therefore Milne boost invariant (by the construction of $g_{\mu\nu}, v_A^\mu, n_\mu, h^{\mu\nu}$), Weyl invariant (by the use of the compensator $\tau$), and $U(1)$ invariant (by the use of the compensator $\theta$).

The other equivalent way to construct the effective action, which is the one we will pursue, is to take Milne boost invariants and perform Weyl and $U(1)$ transformations with parameters $\tau$ and $\theta$, respectively. This will evidently give the same answer (up to the sign of the fields $\tau$ and $\theta$, which will be opposite to the one in the first method, and therefore to the convention used in \cite{Komargodski:2011vj}), since the Weyl and $U(1)$ transformations will force the appearance of $\tau$ and $\theta$ in exactly the right form such that if they themselves transform under Weyl and $U(1)$, the whole expression would be invariant.

Using the boost invariant scalars, we will get the effective action for $\tau$ and $\theta$ after restricting to flat geometry \eqref{eq:flat_geometry}. Note, that the flat geometry restriction is respected by a combination of a Milne boost and a $U(1)$ transformation. Under this combination, $\theta$ transforms in the same way as it transformed under non-relativistic boost transformations in the previous section \eqref{eq:theta_boost_trans}. The reason is that under the appropriate Milne boost transformation parametrized by $\psi_\mu = (0,\vec{u})$ at flat geometry, $A_\mu$ transforms as in \eqref{eq:A_milne_trans}, which in this case takes the form $A_\mu \rightarrow A_\mu + \partial_\mu \left( {-\frac{1}{2}\vec{u}^2t + \vec{u}\cdot\vec{x}} \right) $. In order to compensate for this transformation, the $U(1)$ transformation should be (again, at flat geometry) $\Lambda = \frac{1}{2}\vec{u}^2t - \vec{u}\cdot\vec{x}$, which produces the correct boost transformation of $\theta$.
Note also that the flat geometry restriction is respected by an additional combination of Milne boosts, $U(1)$ transformations and anisotropic Weyl transformations, which corresponds to the special conformal transformation \eqref{eq:SpecialConformalTrans} (see \cite{Jensen:2014aia}). Therefore the effective action obtained using this geometric method necessarily corresponds to the full Schr\"{o}dinger invariant case.

We wish to list the invariant scalars to leading order in derivatives. Note that derivative counting should be done after restricting to flat geometry. Consider the Milne boost invariant metric and vector \eqref{eq:milne_invariant}. The simplest geometrical term is the cosmological constant term (constant up to the $\sqrt{\operatorname{det}(\gamma_{\mu\nu})}$ factor, where $\gamma_{\mu\nu} \equiv h_{\mu\nu}+n_\mu n_\nu$, that contributes the $\tau$ dependence and ensures Weyl invariance), which matches the non-dynamical term $e^{(z+d)\tau}$ \eqref{L0} discussed previously.

The next simplest scalar one can build is $g_{\mu\nu}v_A^\mu v_A^\nu$. We perform Weyl transformation and $U(1)$ transformation and then restrict to flat geometry to get an expression in terms of  $\tau$ and $\theta$. Counting derivatives naively in this expression, $g$ may contribute one derivative and $v_A$ may contribute one space derivative, so we might be lead to think that in total we have 3 derivatives. However, since $h^{\mu\nu}n_{\mu} = 0$, the full expression, when restricted to flat geometry, has only one time derivative or 2 space derivatives. Written explicitly, we get:
\begin{equation}\label{eq:FirstGeometricInvariantTerm}
g_{\mu\nu}v_A^\mu v_A^\nu \rightarrow e^{-2\tau}\left( {2\partial_t\theta -\left(\partial_i \theta\right)^2 } \right) \ .
\end{equation}
The other term at this order in derivatives is the spatial Ricci scalar $\tilde{R}$ corresponding to the standard Levi-Civita connection of the metric induced on the space foliation.\footnote{Since in this context we are considering NC geometries which are conformally flat, i.e.\ $n=e^{2\tau} dt$, we can safely assume that $n_\mu$ satisfies the Frobenius condition and therefore induces a foliation of the spacetime manifold into equal time slices. See \cite{Arav:2016xjc} for further discussion.} 
Since the metric induced on the foliation is invariant under Milne boosts (i.e.\ $g_{\mu\nu}u^\mu w^\nu$ is boost invariant for any space tangent vectors $u^\mu, w^\mu$), $\tilde{R}$ is boost invariant as well.
It is also gauge invariant, and therefore it depends only on $\tau$. By Weyl transformation we get
\begin{equation}\label{eq:SecondGeometricInvariantTerm}
\tilde{R} \rightarrow e^{-2\tau} \left( { -2(d-1)\partial_i^2\tau - d(d-1)\left(\partial_i\tau\right)^2 } \right) \ .
\end{equation}
There are more terms, e.g. $a_\mu a^\mu$ where $a_\mu \equiv - \mathcal{L}_v n_\mu$ (see e.g. \cite{Arav:2016xjc}), but when we restrict to conformally flat background and consider integration by parts, they are all equivalent to the terms above. Therefore, these two expressions complete the list of terms up to one time derivative or two space derivatives (i.e.\ the same order in $z=2$ Lifshitz scaling counting). From the last two expressions, \eqref{eq:FirstGeometricInvariantTerm} and \eqref{eq:SecondGeometricInvariantTerm}, we construct the leading order Lagrangian for the complex field $\phi$ \eqref{L1}.

For higher orders in derivatives, one can define a boost invariant affine connection from the structures $n_\mu, h^{\mu\nu}, v_A^\mu, g_{\mu\nu}$, as well as the corresponding Riemann tensor $R^{\lambda}{}_{\mu\sigma\nu}$ and Ricci tensor $R_{\mu\nu} \equiv R^{\sigma}{}_{\mu\sigma\nu}$ (see \cite{Jensen:2014aia}). One can then obtain various higher derivative boost invariant scalars, such as $ R_{\mu\nu} h^{\mu\nu} $ and $ R_{\mu\nu} v_A^\mu v_A^\nu $, as well as purely spatial ones such as $ \tilde{R}^2 $ and $ a^4 $. As before, by performing Weyl and gauge transformations on a linear combination of these scalars and restricting to flat geometry, one can construct the higher order Lagrangian \eqref{L2} (as well as other higher derivatives terms, which do not affect the conclusions of the analysis in the next subsection).

\subsection{Spectrum Analysis}\label{subsec:BrokenU1Spectrum}
In the following we will analyze the spectrum of the low energy theory of $\tau$ and $\theta$. We use the notation $\phi_{\Delta,m} = e^{\Delta\tau+im\theta}$ for a complex scalar field of dimension $\Delta$ and mass $m$, and $\phi_k$ as a shorthand for $\phi_{\Delta_k,m_k}$. The leading order boost, $U(1)$ and scale invariant Lagrangian which we derived in the previous sections reads
\begin{equation}
\label{eq:TotalLagrangianForSpectrumAnalysis}
\begin{split}
L & = \Lambda \phi_0^* \phi_0 + A \left[ {\frac{i}{2}\left( { \phi_1^*\partial_t\phi_1 - \partial_t\phi_1^*\phi_1 } \right) - \frac{1}{2m_1}\partial_i\phi_1^*\partial_i\phi_1} \right] \\
& + B \left[ { \left( { i\partial_t\phi_2^* - \frac{1}{2m_2}\partial_i^2\phi_2^* } \right) \left( { -i\partial_t\phi_2 -\frac{1}{2m_2}\partial_i^2\phi_2 } \right) } \right] \ ,
\end{split}
\end{equation}
where the dimensions $\Delta_i$ are fixed by scale invariance to the following values:
\begin{equation}
\Delta_0 = \frac{d+2}{2}, \qquad \Delta_1 = \frac{d}{2}, \qquad \Delta_2 = \frac{d-2}{2}.
\end{equation}
Note that this is not the most general expression one could build up to this order in derivatives -- we could take a linear combination of similar terms using fields $\phi_{\Delta,m}$ with various values of $\Delta$ and $m$, while compensating for the dimension by multiplying each by an appropriate exponent of $\tau$ (as in \eqref{L2}). However, for the cosmological constant term (the $\Lambda$ term) and the leading term in derivatives (the $A$ term), one can always rewrite these terms in the form given in \eqref{eq:TotalLagrangianForSpectrumAnalysis} (using just $\phi_0$ and $\phi_1$ with no extra exponents of $\tau$), with an appropriate choice of the parameters $\Lambda$, $A$ and $m_1$. For the subleading term in derivatives (the $B$ term), this is not true in general, but since we will be mainly interested in the leading contributions of the $\Lambda$ and $A$ terms, we will assume the form given in \eqref{eq:TotalLagrangianForSpectrumAnalysis} as an example for the contribution of subleading terms in derivatives. We therefore use the Lagrangian \eqref{eq:TotalLagrangianForSpectrumAnalysis}, for which all $\tau$ dependence is through the $\phi_0$, $\phi_1$ and $\phi_2$ fields. It is also important to note that the subleading term in \eqref{eq:TotalLagrangianForSpectrumAnalysis} is not Schr\"{o}dinger invariant, as mentioned in subsection \ref{subsec:EffActionBrokenU1}. Full Schr\"{o}dinger invariance restricts this subleading term to the form \eqref{L2} with $\Delta_2 = \frac{d}{2} $. However, using such a term instead in \eqref{eq:TotalLagrangianForSpectrumAnalysis} does not change the main results of this subsection.

We derive the equations of motion by varying the action defined by \eqref{eq:TotalLagrangianForSpectrumAnalysis} with respect to $\tau$ and $\theta$, through $\phi_0, \phi_1$ and $\phi_2$. We have $\delta S = \frac{\delta S}{\delta \phi_i} \delta\phi_i + c.c.$, where:
\begin{equation} \label{eq:spectrum_phi_variations}
\begin{split}
& \frac{1}{\Lambda}\frac{\delta S}{\delta \phi_0} = \phi_0^* \ ,\\
& \frac{1}{A}\frac{\delta S}{\delta \phi_1} = -i\partial_t \phi_1^* + \frac{1}{2m_1}\partial_i^2\phi_1^* \ ,\\
& \frac{1}{B}\frac{\delta S}{\delta \phi_2} = -\partial_t^2 \phi_2^* - \frac{i}{m_2}\partial_t\partial_i^2\phi_2^* + \frac{1}{4m_2^2}\partial_i^2\partial_i^2\phi_2^* \  ,
\end{split}
\end{equation}
and their complex conjugates.

In terms of $\tau$ and $\theta$ we have
$\delta \phi_{\Delta,m} = \left( { \Delta\delta\tau + im\delta\theta } \right) \phi_{\Delta,m}$, and therefore:
\begin{equation}
\begin{split}
\delta S = \frac{\delta S}{\delta \phi_0} \phi_0\left( { \Delta_0\delta\tau + im_0\delta\theta } \right) &+ \frac{\delta S}{\delta \phi_1} \phi_1\left( { \Delta_1\delta\tau + im_1\delta\theta } \right)\\
&+ \frac{\delta S}{\delta \phi_2} \phi_2 \left( { \Delta_2\delta\tau +im_2\delta\theta } \right) + \text{c.c.} \ .
\end{split}
\end{equation}
Separating the variation by $\tau$ and $\theta$ we have 
\begin{equation}
\begin{split}
\frac{\delta S}{\delta \tau} & \propto \Delta_0 \Re{\frac{\delta S}{\delta \phi_0} \phi_0}  + \Delta_1 \Re{\frac{\delta S}{\delta \phi_1} \phi_1} + \Delta_2 \Re{\frac{\delta S}{\delta \phi_2} \phi_2} \ , \\
\frac{\delta S}{\delta \theta} & \propto m_1 \Im{\frac{\delta S}{\delta \phi_1} \phi_1} + m_2 \Im{\frac{\delta S}{\delta \phi_2} \phi_2} \ ,
\end{split}
\end{equation}
where $\Re{}$ and $\Im{}$ denote the real and imaginary parts respectively.

We will perturb the equations of motion around fixed values 
\begin{equation}
\tau=\hat{\tau},~~~~\partial_t \theta=-\hat{\mu} \ ,
\end{equation}
where $\hat{\mu}$ is the chemical potential.\footnote{Note that our sign convention for $\theta$ is different from that in \cite{Son:2005rv}.} The particular value of $\hat{\tau}$ can be absorbed by a redefinition of $A$,$B$ and $\Lambda$, so we will take it to be zero. Note however that $v \equiv e^{-\hat{\tau}}$ quantifies the scale of the state we are perturbing about.

We express $\frac{\delta S}{\delta \phi_{1}} \phi_1$ and $\frac{\delta S}{\delta \phi_{2}} \phi_2$ using $\tau$ and $\theta$ (see \eqref{eq:spectrum_phi_variations}):
\begin{equation}
\begin{split}
\frac{1}{A}\frac{\delta S}{\delta \phi_1} \phi_1 & = -i\left( { \partial_t \phi_1^* } \right) \phi_1 + \frac{1}{2m_1} \left( { \partial_i^2\phi_1^* } \right) \phi_1 \\
& = e^{2\Delta_1\tau}\left( { -i\Delta_1\partial_t\tau - m_1\partial_t\theta } \right) + \\
& + \frac{e^{2\Delta_1\tau}}{2m_1}\left( { \Delta_1\partial_i^2\tau -im_1\partial_i^2\theta + \Delta_1^2\left(\partial_i\tau\right)^2 -m_1^2\left(\partial_i\theta\right)^2 - 2im_1\Delta_1\partial_i\tau\partial_i\theta} \right) \\
& \simeq e^{2\Delta_1\tau}\left( { -i\Delta_1\partial_t\tau - m_1\partial_t\theta + \frac{1}{2m_1}\left( { \Delta_1\partial_i^2\tau -im_1\partial_i^2\theta} \right) } \right) \ ,
\end{split}
\end{equation}
where $\simeq$ here denotes keeping only terms that contribute up to linear order in the $\tau$ and $\theta$ perturbation.
Similarly, again up to linear order, we have
\begin{equation}
\begin{split}
\frac{1}{B}\frac{\delta S}{\delta \phi_2} \phi_2 & = -\left(\partial_t^2 \phi_2^*\right)\phi_2 - \frac{i}{m_2}\left(\partial_t\partial_i^2\phi_2^*\right)\phi_2 + \frac{1}{4m_2^2}\left(\partial_i^2\partial_i^2\phi_2^*\right)\phi_2 = \\
& \simeq e^{2\Delta_2\tau} \left( { -\Delta_2\partial_t^2\tau + im_2\partial_t^2\theta + m_2^2 \left( \partial_t\theta \right)^2} + 2i\Delta_2 m_2 \partial_t \theta \partial_t \tau  \right) - \\
& - \frac{i}{m_2}e^{2\Delta_2\tau}\left( {\Delta_2\partial_t\partial_i^2\tau - im_2\partial_t\partial_i^2\theta - m_2^2\partial_i^2\theta\partial_t\theta - i\Delta_2 m_2 \partial_i^2\tau \partial_t\theta} \right) + \\
& + \frac{1}{4m_2^2} e^{2\Delta_2\tau} \left( { \Delta_2\partial_i^2\partial_i^2\tau - im_2\partial_i^2\partial_i^2\theta } \right) \ .
\end{split}
\end{equation}
Combining the last two equations, the variation of the action with respect to $\tau$ and $\theta$ is given by the following expressions up to linear order in the perturbation:

\begin{equation}
\begin{split} 
\label{eq:variation}
\frac{\delta S}{\delta \tau} & \propto \Lambda\Delta_0e^{2\Delta_0\tau} + A \Delta_1 e^{2\Delta_1\tau} \left( { -m_1\partial_t\theta + \frac{\Delta_1}{2m_1}\partial_i^2\tau } \right) 
\\
& + B\Delta_2e^{2\Delta_2\tau} \left( { -\Delta_2\partial_t^2\tau +m_2^2 \left( \partial_t \theta \right)^2 -\partial_t\partial_i^2\theta - \Delta_2 \partial_i^2\tau \partial_t\theta + \frac{\Delta_2}{4m_2^2}\partial_i^2\partial_i^2\tau } \right), \\ 
\frac{\delta S}{\delta \theta} & \propto Am_1e^{2\Delta_1\tau} \left( { -\Delta_1\partial_t\tau - \frac{1}{2}\partial_i^2\theta } \right) 
\\
& + Bm_2 e^{2\Delta_2\tau} \left( { m_2\partial_t^2\theta + 2\Delta_2 m_2 \partial_t\theta \partial_t\tau -\frac{\Delta_2}{m_2}\partial_t\partial_i^2\tau + m_2\partial_i^2\theta\partial_t\theta - \frac{1}{4m_2}\partial_i^2\partial_i^2\theta } \right).
\end{split}
\end{equation}

The zeroth order equation requires 
\begin{equation}
\Lambda\Delta_0 = -Am_1\Delta_1\hat{\mu} - Bm_2^2\Delta_2\hat{\mu}^2 \ ,
\label{A}
\end{equation}
and we see that the static potential value (cosmological constant) $\Lambda$ corresponds to a non-zero chemical potential $\hat{\mu}$ (although a non-zero chemical potential is possible even when $\Lambda$ vanishes). 

Consider first the case of a non-zero chemical potential $\hat{\mu}$. We get the following linearized equations of motion to leading order in momentum $\vec{k}$ and energy $\omega$,
\begin{equation}
 \left( \begin{array}{cc}
 -A\Delta_1 m_1 \omega - 2B\Delta_2 m_2^2 \hat{\mu} \omega &
 2\Lambda\Delta_0^2 + 2A\Delta_1^2 m_1\hat{\mu} + 2B\Delta_2^2m_2^2\hat{\mu}^2 \\
 \frac{1}{2}A m_1 \vec{k}^2 - B m_2^2 \omega^2 + B m_2^2 \hat{\mu} \vec{k}^2 &
 A\Delta_1 m_1 \omega + 2B \Delta_2 m_2^2 \hat{\mu}\omega
 \end{array} \right)
 \left(\begin{array}{c}
 i \delta\tilde{\theta} \\
 \delta\tilde{\tau}
 \end{array}\right) = 0,
\end{equation}
where $\delta\tilde{\tau}$ denotes the Fourier transform of the $\tau$ perturbation around $0$, and $\delta\tilde{\theta}$ denotes the Fourier transform of the $\theta$ perturbation around $-\hat{\mu} t$. 
From these equations we obtain the following dispersion relation:
\begin{equation}\label{eq:SpectrumWithChemicalPotential}
\omega^2 = \frac{1}{\Delta_1} \hat{\mu} \vec{k}^2 .
\end{equation} 
We can see that in this case, in the limit $k \rightarrow 0$, the dispersion relation is linear $\omega \sim k$ and $\delta\tilde\tau \sim \omega \delta\tilde\theta$.
This is consistent with the analysis in \cite{Son:2005rv}.
Note, that since $\delta\tilde\tau \sim \omega \delta\tilde\theta$, the perturbation is mainly in $\theta$, which may also justify ignoring the $\tau$ contribution in the leading order superfluid effective field theory as was implicitly done in \cite{Son:2005rv}. Also note that the stability of the modes in the $k\to 0$ limit requires $\hat{\mu} > 0$.

The result \eqref{eq:SpectrumWithChemicalPotential} implies the speed of sound $ v_s \equiv \sqrt{\frac{1}{\Delta_1}\hat{\mu}} = \sqrt{\frac{2}{d}\hat{\mu}} $. This can be easily understood from dimensional analysis considerations: In the presence of a $z=2$ Lifshitz scale invariance, we can expect the relation between the chemical potential $\hat{\mu}$ and the conserved particle number (or mass) density $\rho$ to be of the form $ \hat{\mu} = C \rho^{2/d} $, where $C$ is some dimensionless parameter. From standard thermodynamic relations, the speed of sound will be given by:
\begin{equation}
v_s^2 = \frac{\partial P}{\partial \rho} = \rho \frac{\partial\hat{\mu}}{\partial\rho} = \frac{2}{d} \hat{\mu} .
\end{equation}

Consider next the case $\Lambda=\hat{\mu}=0$. 
The linearized equations are given by:
\begin{equation}
\label{harmonic} 
 \left( \begin{array}{cc}
 A\Delta_1 m_1 \omega - B\Delta_2\omega\vec{k}^2 & A \frac{\Delta_1^2}{2m_1}\vec{k}^2 - B\Delta_2^2\omega^2 - B\frac{\Delta_2^2}{4m_2^2}\vec{k}^4\\
 \frac{1}{2}Am_1\vec{k}^2-Bm_2^2\omega^2-\frac{1}{4}B\vec{k}^4 & A\Delta_1m_1\omega - B\Delta_2\omega\vec{k}^2 \end{array}
\right) \left( \begin{array}{c}
         i\delta\tilde{\theta} \\
         \delta\tilde{\tau}
        \end{array}
 \right) = 0 \ . 
\end{equation}
When $A\neq 0$ we have at leading order for small values of $\vec{k}$ (and therefore $\omega$) the non-relativistic dispersion relation
\begin{equation}
\omega = \frac{\vec{k}^2}{2m_1}, ~~~ \delta\tilde{\tau} = -i\frac{m_1}{\Delta_1}\delta\tilde{\theta} \ .
\label{dis}
\end{equation}
Thus, we find one gapless mode at large length scales compared to the breaking scales. The corrections can be computed to give $\omega^2 = \alpha k^4 + \beta k^6 + \ldots\, $, where the coefficients $\alpha,\beta,\ldots$ are determined from expanding the determinant of \eqref{harmonic} to growing orders in $\vec{k}^2$.
Note that the limit $m_1\rightarrow 0$ takes us from the broken $U(1)$ to the unbroken $U(1)$ case. The dispersion relation \eqref{dis} blows up and we are left with no propagating mode. 

In addition to the gapless mode, we have also a gapped mode as can be seen by setting $\vec{k}=0$ in \eqref{harmonic} and we get
 \begin{equation}
 \omega^2 = \frac{A^2\Delta_1^2m_1^2}{B^2\Delta_2^2m_2^2} > 0, ~~~ \delta\tilde{\tau} = -i\frac{m_2}{\Delta_2}\delta\tilde{\theta} \ .
\end{equation}

Finally, note that we can obtain the case of $U(1)$ SSB without scale invariance from the above equation \eqref{eq:variation}. Take $\tau$ to be a constant rather than  a dynamical field, and to first order the variation with respect to $\theta$ gives
\begin{equation}
 \frac{\delta S}{\delta \theta} \propto Am_1e^{2\Delta_1\tau} \left( { - \frac{1}{2}\partial_i^2\theta } \right) + Bm_2 e^{2\Delta_2\tau} \left( { m_2\partial_t^2\theta  - m_2\hat{\mu}\partial_i^2\theta} \right) \ ,
\end{equation}
which leads to a linear dispersion relation. Note that the cosmological constant term does not contribute to this result.

\section{On a Non-Relativistic a-theorem and the Frobenius condition} \label{sec:atheorem}

The relativistic dilaton effective action was valuable for the proof of the a-theorem \cite{Komargodski:2011vj} in $3+1$ dimensions, i.e.\ the coefficient of the A-type conformal anomaly $a$ satisfies $a_{\text{IR}} < a_{\text{UV}}$. In the following we will make a few comments on the non-relativistic Galilean case and the feasibility of using similar arguments to prove an RG flow theorem in case such a theorem indeed holds. 

In \cite{Komargodski:2011vj}, the RG flow of a generic relativistic field theory from a UV to an IR fixed point was studied by weakly coupling the theory to a dilaton as a conformal compensator, and matching the conformal anomalies between the UV and IR theories. In particular, the A-type anomaly of the theory contributes to the effective action of the dilaton in the IR, and therefore the S-matrix of dilaton scattering. The a-theorem then follows from unitarity of the theory. 

In the case of non-relativistic field theories,  invariance under a Lifshitz scale symmetry implies the following Ward identity for the stress-energy tensor: 
\begin{equation}\label{eq:LifWardIdent}
D \equiv T^{\mu\nu} h_{\mu\nu} - z T^{\mu\nu} n_\mu n_\nu = 0 \ ,
\end{equation}
which is just a generalized version of the conformal tracelessness condition.
However, similarly to the relativistic case, the scale symmetry can be violated due to quantum anomalies (analogous to the conformal anomalies) \cite{Baggio:2011ha,Griffin:2011xs,Arav:2014goa,Arav:2016xjc,Auzzi:2015fgg,Auzzi:2016lrq,Arav:2016akx}. The expectation value of the stress-energy tensor on a curved spacetime manifold then no longer satisfies identity \eqref{eq:LifWardIdent}. It instead acquires an anomalous contribution:\footnote{Such non-relativistic scale anomalies also appear as contact terms in correlation functions of the flat space theory involving the operator $D$  \cite{Arav:2016akx}.}
\begin{equation}
\left\langle D \right\rangle \equiv \left\langle T^{\mu\nu} \right\rangle h_{\mu\nu} - z \left\langle T^{\mu\nu} \right\rangle n_\mu n_\nu = \mathcal{A} \ ,
\end{equation}
where $\mathcal{A}$ is a local functional of the backgrounds fields, and the infinitesimal (anisotropic) Weyl transformation of the effective action is given by:
\begin{equation}\label{eq:EffectiveActionWeylTrans}
\delta_\sigma S_\text{eff} = \int \sqrt{\gamma}\,\sigma \mathcal{A} \ .
\end{equation}

For Galilean invariant theories in $d+1$ dimensions, it has been suggested in \cite{Jensen:2014aia} that these Lifshitz anomalies correspond to conformal anomalies of relativistic field theories in $d+2$ dimensions defined on a manifold with a null isometry, via a null reduction procedure. In particular, this suggests a possible A-type anomaly in these Galilean theories, which corresponds to the Euler density anomaly term of the $d+2$ dimensional relativistic theory. In \cite{Arav:2016xjc}, this possibility was confirmed for a $2+1$ dimensional Galilean theory via a cohomological analysis of the Wess-Zumino consistency condition (an explicit expression for this A-type anomaly is given in equation (5.13) of \cite{Arav:2016xjc}). However, it was observed that this A-type anomaly exists only when one assumes the Frobenius condition is violated by the curved spacetime NC structure, i.e.\ when the 1-form $n_\mu$ does not satisfy:
\begin{equation}\label{eq:FrobeniusCondition}
n \wedge dn = 0 \ ,
\end{equation}
and therefore does not induce a foliation of the spacetime manifold into equal-time slices. When such a foliation structure exists, this A-type anomaly term becomes cohomologically trivial, and can be removed by adding an appropriate local counter-term to the effective action, of the form:
\begin{equation}\label{eq:LCounterterm}
L_\text{c.t.} = \frac{1}{2} a^\mu \partial_\mu \left( a^2 \right) + \frac{3}{8} a^4,
\end{equation}
where $a_\mu \equiv - \mathcal{L}_v n_\mu $ is the acceleration associated with $n_\mu$ (see equation (5.14) in \cite{Arav:2016xjc}).

The existence of the A-type scale anomaly in the Galilean case suggests it may be possible to follow a similar argument to the one given in \cite{Komargodski:2011vj} to prove an a-theorem for Galilean field theories. Consider a theory which is invariant under the Galilean group, that flows between a UV and an IR $z=2$ Lifshitz fixed points. This theory can be coupled to a non-relativistic dilaton field $\tau$ by multiplying any dimensionful parameter by an appropriate exponent of $\tau$ compensating for its dimension, thereby rendering the theory scale invariant. We can also add a kinetic term for the dilaton, however as we have seen in previous sections in order to have a boost invariant, dynamic term one has to involve the $U(1)$ particle number Goldstone mode $\theta$.\footnote{Alternatively one can use the Galilean boosts modes $v_i$ (as considered in appendix \ref{boostsAppendix}). We will focus here on the $U(1)$ mode, but the other option isn't significantly different.} We can then choose this kinetic term to be of the form \eqref{L1}\footnote{Alternatively we could discuss an RG flow triggered by some operator acquiring a VEV that spontaneously breaks both scale invariance and the $U(1)$ particle number symmetry, leading to a dynamic NGB effective action of the form discussed in previous sections.} (with an arbitrary value of $m$). The coupling of the matter to the dilaton can be made arbitrarily weak by taking the coefficient of this term $A$ to be much larger than all other dimensionful parameters.

Similarly to the way equation (3.2) in \cite{Komargodski:2011vj} was derived, we can write an IR effective action for the non-relativistic Galilean theory coupled to the dilaton.
As in the relativistic case, the IR effective action will have a contribution $\text{NR}_{\text{IR}}$ from the non-relativistic Lifshitz invariant Galilean field theory (that replaces the relativistic CFT) in the infrared. It will have a contribution from the invariant terms, $L_\text{dilaton}$, corresponding to the local effective action of the dilaton as discussed in the previous sections, which is of the general form \eqref{eq:TotalLagrangianForSpectrumAnalysis} (with possibly more terms of the 2 time or 4 space derivative order or higher). And finally, it will have a contribution from the A-type anomaly of the theory. This contribution can be calculated by replacing the Weyl parameter $\sigma$ in \eqref{eq:EffectiveActionWeylTrans} by $\tau$, substituting into $\mathcal{A}$ the expression for the A-type anomaly (given in \cite{Arav:2016xjc} for $2+1$ dimensions) evaluated on a NC background which is Weyl transformed\footnote{Since we are assuming the $U(1)$ symmetry is not anomalous, the anomalous contribution is gauge invariant, and so will not depend on $\theta$ if a gauge transformation is performed with $\theta$ as the parameter.} with $\tau$ as the parameter, solving the resulting equation and restricting to flat space (see e.g.\ \cite{Schwimmer:2010za}). Alternatively, it can be obtained from the conformal A-type anomaly contribution in $d+2$ dimensions via a null reduction. The result for $2+1$ dimensions is similar to the $3+1$ dimensional relativistic case, and given by:
\begin{equation}
\begin{split}
 S_{\text{IR}} &= \text{NR}_{\text{IR}} \\
 & + \int{dtd^2x \left( { L_\text{dilaton} + \left( a_{\text{UV}} - a_{\text{IR}} \right) \left( { -4\left(\partial_i\tau\right)^2 \partial_i^2\tau - 2\left(\partial_i\tau\right)^4 } \right)  } \right) } ,
\end{split}
\label{S}
\end{equation}
where $a_{\text{UV}}$ and $a_{\text{IR}}$ are the coefficients of the non-relativistic A-type anomaly in the UV and the IR theories respectively (including the dilaton contribution). Note that the A-type anomaly contribution has no time derivatives. This is a consequence of the fact that it is a $U(1)$ singlet.

Similarly to the relativistic case, the dilaton action $L_\text{dilaton}$ has additional higher order scale invariant terms, whose couplings are non-universal and cannot be fixed. However, there is an important difference: In the relativistic case, the 4 derivatives terms in $L_\text{dilaton}$ are distinguishable from the contribution of the A-type anomaly, that is the anomaly contribution term cannot be reproduced by a linear combination of the allowed invariant terms in $L_\text{dilaton}$. This is not true for the Galilean case in $2+1$ dimensions, since the anomaly contribution term in \eqref{S} can also be obtained from a local contribution of the form \eqref{eq:LCounterterm} to $L_\text{dilaton}$. This observation can be understood as a consequence of the Frobenius condition: In order to obtain the dilaton effective action in flat space, one naturally works with a conformally flat background, which necessarily satisfies the Frobenius condition \eqref{eq:FrobeniusCondition} and therefore has a foliation structure. On such a background, as discussed in \cite{Arav:2016xjc}, there is no A-type anomaly, as it becomes cohomologically trivial. In order to be able to distinguish the anomaly contribution from non-universal contributions to $L_\text{dilaton}$, one would have to instead look at a field theory defined on a curved NC background that violates the Frobenius condition. 

It is also important to note here the role played by the special conformal transformation. In the relativistic case, the contribution of the A-type anomaly to the dilaton effective action is invariant under global scaling transformations. It is not, however, invariant under special conformal transformations, and it is this property that prevents it from being included in $L_\text{dilaton}$ from the point of view of the flat space theory. In the non-relativistic case, imposing full Schr\"{o}dinger invariance does not have the same consequence, as the (purely spatial) contribution of the A-type anomaly is in fact invariant under the Schr\"{o}dinger special conformal transformations.

Another difference compared to the relativistic case is that these higher order terms in $L_\text{dilaton}$ may contribute to the dilaton 2 to 2 scattering, in contrast to the relativistic CFT case where the higher order terms didn't contribute at leading order. 
There are two other notable differences that have already been mentioned: 
The first is that, while the non-relativistic anomaly term is a $U(1)$ singlet \eqref{S}, the dilaton effective action $L_\text{dilaton}$ in the non-relativistic case involves the $U(1)$ particle number Goldstone mode. 
Second, in the relativistic CFT case where the RG flow is triggered by conformal SSB, a cosmological constant leads to a gapped mode in contradiction to the Goldstone theorem, and is not allowed in the SSB effective action. On the other hand, in the non-relativistic case as we showed above, a cosmological constant is allowed and simply leads to a chemical potential.

These differences, and especially the first one, namely the fact that in a conformally flat background the A-type anomaly contribution is indistinguishable from that of a trivial term, seem to suggest that it is not straightforward to generalize the  proof of the a-theorem to the Galilean case. 

\section{Summary and Outlook}
We studied the mechanism of spontaneous symmetry breaking of scale invariance in non-relativistic field theories that possess Galilean boost invariance. We showed that there is a dynamical gapless mode only if the $U(1)$ particle number symmetry or the Galilean boost symmetries are spontaneously broken too. The dispersion relation of the gapless mode  depends on the breaking pattern and on the chemical potential. It is quadratic in the case of spontaneously broken particle number symmetry and zero chemical potential, and linear in all other cases. We constructed the leading terms in the dilaton effective action in two ways: First by using symmetry arguments and second by employing the Newton-Cartan geometrical structure.

The effective action of the dilaton in relativistic field theories encodes the information of trace anomalies. We considered the question whether and how the non-relativistic scale anomalies are encoded in the non-relativistic dilaton effective action. We found that there is a major difference between the relativistic and non-relativistic cases. The construction of the dilaton effective action in flat space requires working with a conformally flat background. However, such a background satisfies the Frobenius condition and therefore implies that the A-type anomaly is cohomologically trivial \cite{Arav:2016xjc}. Thus, in contrast to the relativistic case, in order to distinguish the anomaly contribution from non-universal contributions to the dilaton action, one has to consider a curved NC background that violates the Frobenius condition. The study of such field theories and their consistency is an important challenge that we leave for the future. This is likely to shed light on the question whether non-relativistic RG theorems analogous to the relativistic ones exist. Another interesting topic left for the future, which is potentially linked to the structure of the RG flows, is the role of the special conformal generator in the symmetry algebra of non-relativistic field theories. This is the non-relativistic version of scale versus conformal invariance of relativistic field theories.

\section*{Acknowledgments}
We would like to thank S. Chapman, C. Eling, J. Gomis, C. Hoyos, Z. Komargodski and S. Theisen for valuable discussions and comments. This work is supported in part by the I-CORE program of Planning and Budgeting Committee (grant number 1937/12), the US-Israel Binational Science Foundation, GIF and the ISF Center of Excellence. I.A is grateful for the support of the Alexander Zaks Scholarship.

\appendix

\section{The Case of Broken Boosts With Unbroken $U(1)$ Symmetry} \label{boostsAppendix}

In this appendix we derive the first few terms in a derivative expansion of the effective action of the non-relativistic NG bosons for the case of broken Lifshitz scale and Galilean boost symmetries, but unbroken $U(1)$ symmetry. We also analyze the corresponding NG modes spectrum.

\subsection{Symmetries} \label{boostAppendixSymmetries}

We consider the case where Galilean boost symmetries are spontaneously broken but particle number symmetry is conserved. In this case, we expect to find NG fields $v_i$ corresponding to the broken Galilean boosts generators, and a NG field $\tau$ corresponding to the broken scale generator as usual.
The transformations of these fields under non-relativistic scale transformation are clearly given by:
\begin{equation}
 \tau \rightarrow \tau + \sigma, ~~~ v_i \rightarrow e^{\sigma} v_i,
\end{equation}
whereas their transformations under Galilean boosts are given by:
\begin{equation}
 \tau \rightarrow \tau, ~~~ v_i \rightarrow v_i + u_i
\end{equation}
Using the field $v_i$, one can build the covariant differential operator
\begin{equation}
 \partial_t + v_j \partial_j,
\end{equation}
which is dimensionful but boost invariant. By acting with this operator on various covariant objects, we can obtain the possible effective action terms which include time derivatives. Furthermore, we may use simple space derivatives at will, as long as we keep rotation invariance.
Finally, we compensate for the dimension using an exponent of $\tau$ of the corresponding dimension.

Listing the allowed terms up to two derivatives, including only derivatives of $v_i$, we have:
\begin{equation}
 e^{-6\tau}\left( \partial_t v_i + v_j\partial_j v_i \right) ^ 2, ~ e^{-4\tau}\left( \partial_i v_i \right)^2, ~ e^{-4\tau}\left( \partial_i v_j \right) ^ 2, ~ e^{-4\tau} \partial_i v_j \partial_j v_i \ ,
\end{equation}
including only derivatives of $\tau$ we have:
\begin{equation}
 e^{-2\tau} \left( \partial_t \tau + v_i \partial_i \tau \right), ~ e^{-4\tau} \left(\partial_t \tau + v_i \partial_i \tau \right)^2, ~ e^{-2\tau} \left( \partial_i \tau \right) ^ 2 \ ,
\end{equation}
and finally including both $\tau$ and $v_i$ derivatives we also have:
\begin{equation}
 e^{-4\tau}\left( \partial_t \tau + v_i \partial_i \tau \right) \partial_j v_j \ .
\end{equation}

A cosmological-constant-like term $\Lambda$ is also allowed by symmetries, but is forbidden in the NGB effective action in this case since it induces a gap. In parity violating theories in $2+1$ dimensions one may also have terms such as $\epsilon_{ij} \partial_i v_j$, but here we restrict ourselves to parity invariant theories (or to $d>2$).

Overall, the Lagrangian density takes the following form up to second order in derivatives:
\begin{equation}\label{boostAppendixEffectiveAction}
\begin{aligned}
 L  = e^{\left( d+2 \right)\tau} & \left[ A e^{-6\tau}\left( \partial_t v_i + v_j\partial_j v_i \right) ^ 2 + B e^{-4\tau}\left( \partial_i v_i \right)^2 + C e^{-4\tau}\left( \partial_i v_j \right) ^ 2 + D e^{-4\tau} \partial_i v_j \partial_j v_i  \right. \\ 
 & \left. + E e^{-2\tau} \left( \partial_t \tau + v_i \partial_i \tau \right) + F e^{-4\tau} \left(\partial_t \tau + v_i \partial_i \tau \right)^2 + G e^{-2\tau} \left( \partial_i \tau \right) ^ 2 \right. \\
 & \left. + H e^{-4\tau}\left( \partial_t \tau + v_i \partial_i \tau \right) \partial_j v_j \right] \ .
\end{aligned}
\end{equation}

For the purpose of studying the linearized equations of motion, it is sufficient to expand this effective action up to second order in the field perturbations of $\tau$ around some value $\hat{\tau}$ and $v_i$ around $0$. The particular value of $\hat{\tau}$ can be absorbed by a redefinition of the coefficients in \eqref{boostAppendixEffectiveAction}, and will be set to $0$.  Note that, since all terms in the action \eqref{boostAppendixEffectiveAction} are of second order in the field perturbations (except for the one time derivative term $ e^{-2\tau}\partial_t\tau $ which is a total derivative), the exponentials do not contribute at this order. Up to second order in derivatives and field perturbations (and taking into account integration by parts), one is left with the following independent terms:
\begin{equation}\label{boostAppendixSecondOrderAction}
 \left( \partial_t v_i \right) ^ 2, ~ \left( \partial_i v_i \right)^2, ~ \left( \partial_i v_j \right) ^ 2, ~ v_i \partial_i \tau, ~ \left(\partial_t \tau \right)^2, ~ \left( \partial_i \tau \right) ^ 2, ~ \partial_t \tau \partial_i v_i \ .
\end{equation}
Note that out of these leading order terms, only $v_i \partial_i \tau$ and $\partial_t \tau \partial_i v_i$ mix the $v_i$ and $\tau$ perturbations.

We may also consider the case of fully Schr\"{o}dinger invariant theories, which requires in addition invariance under the special conformal transformation. For the field $\tau$, this transformation is given in \eqref{eq:SpecialConformalTrans}, whereas for $\vec{v}$ it is given by:
\begin{equation}
 \vec{v} \left( \vec{x},t \right) \rightarrow \frac{1}{ 1 + \nu t } \left(  \vec{v} \left( \frac{\vec{x}}{1+\nu t} , \frac{t}{1+\nu t} \right) + \nu \vec{x} \right) . \\
\end{equation}
Not all of the terms in \eqref{boostAppendixEffectiveAction} are invariant under this transformation -- this added requirement restricts the possible terms to the following independent combinations:
\begin{equation}
\begin{aligned}
& e^{-4\tau} \left[ \partial_t\tau + v_i \partial_i \tau  + \frac{1}{d} \partial_i v_i \right]^2, ~ e^{-4\tau} \left[ \partial_{(i} v_{j)} - \frac{1}{d} (\partial_k v_k) \delta_{ij} \right] ^2, \\
 & e^{-6\tau} \left[ \partial_t v_i + v_j \partial_j v_i \right]^2, 
 ~ e^{-4\tau} \left[ \partial_t v_i + v_j \partial_j v_i \right] \partial_i \tau, \\
& e^{-4\tau} \left[ \partial_i v_j - \partial_j v_i \right]^2,
 ~ e^{-2\tau} {\left(\partial_i\tau\right)^2 }\ .
\end{aligned}
\end{equation}
In particular, out of the two mixing terms only the two-derivatives one is allowed. As we will see below, the other term can be ruled out by stability considerations as well.

\subsection{Geometry}

Similarly to subsection \ref{subsec:BrokenU1Geometry}, we can also construct the NG boson effective action from geometric considerations. Following a similar path, we consider all possible gauge invariant expressions (since the $U(1)$ particle number symmetry is unbroken in this case), perform Milne boost and Weyl transformations (see \eqref{eq:v_milne_trans}--\eqref{eq:A_milne_trans}, \eqref{eq:Weyl_trans}) and restrict to flat geometry \eqref{eq:flat_geometry}.

To that end, we first define the following space tangent tensors (see e.g. \cite{Arav:2016xjc}):
\begin{align}
(K_S)_{\mu\nu} &\equiv - \frac{1}{2} \mathcal{L}_v h_{\mu\nu},\\
K_S &\equiv (K_S)^\mu_\mu, \\
(K_S^\text{tl})_{\mu\nu} &\equiv (K_S)_{\mu\nu} - \frac{1}{d} K_S h_{\mu\nu}, \\
E_\mu &\equiv  - F_{\mu\nu} v^\nu, \\
B_{\mu\nu} &\equiv h^{\mu'}_\mu h^{\nu'}_\nu F_{\mu'\nu'},
\end{align}
where $F_{\mu\nu}$ is the field strength tensor associated with the gauge field $A_\mu$, as well as the previously mentioned $a_\mu \equiv - \mathcal{L}_v n_\mu $ and the spatial Ricci scalar $\tilde{R}$.

The zero derivative invariant scalar that can be constructed is the cosmological constant term $\Lambda$ mentioned in subsection \ref{boostAppendixSymmetries} (and ruled out since it creates a gap). At first order in derivatives, the only non-vanishing scalar that can be constructed is $K_S$, which is a total derivative and therefore does not contribute to the effective action (again, we consider only parity invariant terms).
The possible independent two derivative scalars, and their contributions to the effective action (after performing Milne boost and Weyl transformations and restricting to flat geometry) are as follows:\footnote{The Milne boost transformation parameter $\psi_\mu$ corresponds here to the boost NG field $v_i$.}
\begin{align}
K_S^2 &\rightarrow e^{-4\tau} \left[ d \left(\partial_t\tau + v_i \partial_i \tau \right) + \partial_i v_i \right]^2, \\
(K_S^\text{tl})_{\mu\nu} (K_S^\text{tl})^{\mu\nu} &\rightarrow e^{-4\tau} \left[ \partial_{(i} v_{j)} - \frac{1}{d} (\partial_k v_k) \delta_{ij} \right] ^2, \\
E^2 &\rightarrow e^{-6\tau} \left[ \partial_t v_i + v_j \partial_j v_i \right]^2, \\
E^\mu a_\mu &\rightarrow -2 e^{-4\tau} \left[ \partial_t v_i + v_j \partial_j v_i \right] \partial_i \tau, \\
B_{\mu\nu} B^{\mu\nu} &\rightarrow e^{-4\tau} \left[ \partial_i v_j - \partial_j v_i \right]^2,
\end{align}
as well as the $\tilde{R}$ given in \eqref{eq:SecondGeometricInvariantTerm} (as before, $a^2$ is related to $\tilde{R}$ via integration by parts after restricting to flat geometry). These terms indeed agree with the terms found in subsection \ref{boostAppendixSymmetries} for the full Schr\"{o}dinger invariant case (which, as mentioned in subsection \ref{subsec:BrokenU1Geometry}, corresponds to this geometric construction).

\subsection{Spectrum Analysis}

In the following, we analyze the spectrum of the low energy theory of the $\tau$ and $v_i$ perturbations. Starting from the scale and boost invariant Lagrangian density \eqref{boostAppendixEffectiveAction} derived in previous subsections, we expand it up to second order in the field perturbations and keep only the independent terms \eqref{boostAppendixSecondOrderAction} to obtain:
\begin{equation}
 L = A \left( \partial_t v_i \right) ^ 2 + B \left( \partial_i v_i \right)^2 + C \left( \partial_i v_j \right) ^ 2 + E v_i \partial_i \tau + F \left(\partial_t \tau \right)^2 + G \left( \partial_i \tau \right) ^ 2  + H \partial_t \tau \partial_i v_i \ .
\end{equation}
This Lagrangian leads to the following linearized equations of motion:
\begin{equation}
 \left( 
 \begin{array}{ccc}
  -2A\omega^2 -2Ck^2 & 0 & 0\\
 0 & -2A\omega^2 -2Bk^2 -2Ck^2 &  -iEk -H\omega k\\
 0& iEk -H\omega k & -2F\omega^2-2G k^2
 \end{array}
\right) \left( \begin{array}{c}
         \delta\tilde{v_\perp} \\
         \delta\tilde{v_\parallel} \\
         \delta\tilde{\tau}
        \end{array}
 \right) = 0 \ ,
\end{equation}
where $\delta\tilde{\tau}$, $\delta\tilde{v_\perp}$ and $\delta\tilde{v_\parallel}$ denote the Fourier transform of the $\tau$ field, the transverse component of the $v$ field ($\partial_i (v_\perp)_i = 0$) and the longitudinal component of the $v$ field ( $\partial_{[i} (v_\parallel)_{j]} = 0$), respectively.

We can see that the transverse $v$ mode follows a linear dispersion relation of the form: 
\begin{equation}
\omega^2 = -\frac{C}{A} k^2,
\end{equation}
which is independent of the longitudinal and $\tau$ modes, while the longitudinal and the $\tau$ modes are mixed. For these mixed modes, one obtains the following equation:
\begin{equation}
 \alpha \omega^4 + \beta\omega^2k^2 + \gamma k^4 + \delta k^2 = 0,
\end{equation}
where:
\begin{equation}
 \begin{aligned}
  \alpha &\equiv 4AF, \\
  \beta &\equiv 4AG + 4BF + 4CF - H^2, \\
  \gamma &\equiv 4CG, \\
  \delta &\equiv -E^2 \ .
 \end{aligned}
\end{equation}
Solving for $\omega^2$, we get the dispersion relation:
\begin{equation}
 \omega^2 = \frac{-\beta k^2 \pm \sqrt{\beta ^2k^4 - 4\alpha \gamma k^4 -4\alpha \delta k^2} }{2\alpha} .
\end{equation}
It can be clearly seen from this relation that when $\delta \neq 0$, $\omega(k)$ becomes complex as $k \to 0$. The stability of the modes therefore requires the coefficient $E$ of the one-derivative mixing term $v_i\partial_i\tau$ to vanish (as mentioned in subsection \ref{boostAppendixSymmetries}, the same condition is required by Schr\"{o}dinger invariance). Imposing this requirement, we get a linear dispersion relation $\omega \propto k$ for the longitudinal $v$ and $\tau$ modes.\footnote{Note that positivity of the Hamiltonian requires that $ \operatorname{sign}(B)=\operatorname{sign}(C)=\operatorname{sign}(G)=-\operatorname{sign}(A)=-\operatorname{sign}(F)$. Given that $E=0$, these conditions also ensure that there are no unstable modes. Also, note that we assume here $\alpha,\beta,\gamma \neq 0$, otherwise analysis of the $k\to 0$ limit requires expanding the action to higher order in derivatives.}
Also note that when $H\neq0$, the $\tau$ and $v_\parallel$ modes are mixed.

We thus conclude that, making the physically reasonable assumption that unstable modes are forbidden, one finds $d+1$ NG modes for the $d$ broken Galilean boost symmetries and one broken scale symmetry as one would expect, all with linear dispersion relations, and with the longitudinal boost and dilaton modes being possibly mixed.

\end{document}